\documentstyle[prl,aps,floats,epsf,color]{revtex}
\begin{document}
\baselineskip=12pt
\def\be{\begin{equation}}
\def\ee{\end{equation}}
\def\bea{\begin{eqnarray}}
\def\eea{\end{eqnarray}}
\def\E{{\rm e}}
\def\bearst{\begin{eqnarray*}}
\def\eearst{\end{eqnarray*}}
\def\peleven{\parbox{11cm}}
\def\peffec{\peight{\bearst\eearst}\hfill\peleven}
\def\pspace{\peight{\bearst\eearst}\hfill}
\def\ptwelve{\parbox{12cm}}
\def\peight{\parbox{8mm}}
\twocolumn[\hsize\textwidth\columnwidth\hsize\csname@twocolumnfalse\endcsname

\title
{ Two-Scale Kirchhoff Theory: Comparison of Experimental
Observations With Theoretical Prediction }

\author
{G. R. Jafari $^a$, P. Kaghazchi$^a$, R. S. Dariani $^c$,\\
A. Iraji zad $^a$, S. M. Mahdavi $^a$, M. Reza Rahimi Tabar
$^{a,b}$ and N. Taghavinia $^a$}

\vskip 1cm

\address
{\it $^a$  Department of Physics, Sharif University of
Technology, P.O. Box 11365-9161, Tehran, Iran \\
$^b$ CNRS UMR 6529, Observatoire de la C$\hat o$te d'Azur, BP
4229, 06304 Nice Cedex 4, France\\
$^c$ Department of Physics, Azzahra University, Tehran, 19834
Iran}
 \maketitle


\begin{abstract}

We introduce a non-perturbative two scale Kirchhoff theory, in the
context of light scattering by a rough surface. This is a two
scale theory which considers the roughness both in the wavelength
scale (small scale) and in the scales much larger than the
wavelength of the incident light (large scale). The theory can
precisely explain the small peaks which appear at certain
scattering angles. These peaks can not be explained by one scale
theories. The theory was assessed by calculating the light
scattering profiles using the Atomic Force Microscope (AFM)
images, as well as surface profilometer scans of a rough surface,
and comparing the results with experiments. The theory is in good
agreement with the experimental
results. 
\end{abstract}
\hspace{.3in}
\newpage
]\section{Introduction}

Wave scattering by rough surfaces has been extensively studied
both analytically and experimentally. For analytical approaches
two methods have been generally considered: rigorous
electromagnetic theory and approximate methods. The Kirchhoff
theory is among the electromagnetic theories and is known as a
"tangent plane theory". This theory is most widely used to
calculate the distribution of the specular and diffuse parts of
the reflected light. The Kirchhoff theory treats any point on a
scattering surface as a part of an infinite plane, parallel to the
local surface tangent. The theory is therefore exact for an
infinite, smooth and planar scatterer, but is approximate for
scatterers that are finite sized, non-planar or for rough
surfaces\cite{Ogilvy}. Due to the computational limitations, most
studies have been done for one dimensional data of the surfaces.
There are only few cases of the analysis of two dimensional
surface data. One and two dimensional exact approaches have been
successfully applied to dielectric, metallic or perfectly
conducting surfaces \cite{Kakuen,Nieto}, deterministic
surfaces\cite{Joyez,Dimenna}, dielectric films on a glass
substrate \cite{Gu} and dielectric films \cite{Ingve,Calvo}. Such
exact calculations have been compared with experimental results
and approximate models \cite{Gu,Jerome}. Also some authors studied
wave scattering from random layers with rough interfaces
\cite{Antoine1,Antoine2}.

The joint probability density functions (PDF) of surface slopes
and heights $P(\partial_x h, h)$, is a key function in the
estimation of the main parameters of wave scattering by a rough
surface \cite{Bass,Ishimaru,Vornovich,Fuks,Bruce}. This is more
obvious in a geometrical optics approach, when the angular
distribution of the scattered power is proportional to the
specular reflecting slope PDF. The slope PDF has also been
introduced in references \cite{Bass,Fuks,Bruce} in the context of
Bragg scattering. They have shown that, the Bragg scattering
results must be averaged by the proper slope PDF of the rough
surface. This is also true for the estimation of the thermal
emission from rough surfaces at small grazing angles
\cite{Fuks,Bruce}.

In the present paper, we introduce a non-perturbative two scale
Kirchhoff theory. The theory is applied to explain the small peaks
observed in the scattering profile of a rough surface, at certain
scattering angles. The theory employs the data obtained from the
rough surface in two different scales. To check the theory we have
measured the scattered light intensity as a function of the
scattering angle, $I(\theta)$, using a setup consisting of a He-Ne
laser (632.8nm), a photo-multiplier tube (PMT) detector and a
computer controlled micro-stepper rotation stage. The resolution
of the micro-stepper was $0.5$ minutes. Alumina sheets were used
as the rough samples. The surface topography of the alumina
samples in small scale $(< 5 \mu m)$ was obtained using an atomic
force microscope (AFM) (Park Scientific Instruments). The images
in small scale were collected in a constant force mode and
digitized into $256 \times 256$ pixels. A commercial standard
pyramidal $Si_{3} N_{4}$ tip was used. A variety of scans, each
with size $L$, where recorded at random locations on the surface.
The large scale $(< 5 mm)$ morphology line scans of the alumina
samples were recorded using a surface profilometer (Taylor
Hobson). Figures (1) and (2) show typical AFM image and surface
profile data with resolutions of about $20 nm$ and $0.25 \mu m$,
respectively.

\section{Non-Perturbative Two Scale Kirchhoff Theory}

The Kirchhoff theory is based on three major assumptions
\cite{Ogilvy}:

 {\bf a)}-  The surface is observed from far field.

 {\bf b)}- The surface
 is regarded as flat, and the optical behavior
 is locally identical at any given point on the surface. Therefore the Fresnel laws
 can be locally applied.

 {\bf c)}- The amplitude of the reflection coefficient, $R_{0}$ is independent
          of the position on the rough surface.

 The field scattered by the rough surface,
 $\psi^{sc}(r)$, is obtained by an integration
 over the mean reference plane $S_{M}$ \cite{Ogilvy}, (the geometry is displayed in figure
 (3))
\begin{eqnarray}\label{3}
&& \psi^{sc}(r)=\frac{ik \exp(ikr)}{4\pi r}\int\int_{s_{M}}
(a\frac{\partial h}{\partial x_{0}}+b \frac{\partial h}{\partial
y_{0}}-c) \cr \nonumber \\
&& \exp{(ik(Ax_{0}+By_{0}+ C h(x_{0},y_{0})))} dx_{0}dy_{0}
\end{eqnarray}
where
\begin{eqnarray}\label{ABC}
A&=&\sin \theta_{1}-\sin \theta_{2} \cos \theta_{3} ,\cr \nonumber \\
B&=&-\sin \theta_{2} \sin \theta_{3} ,\cr \nonumber \\
C&=&-(\cos \theta_{1} +\cos \theta_{2}) ,\cr \nonumber \\
a&=&\sin \theta_{1}(1-R_{0})+\sin \theta_{2}\cos
\theta_{3}(1+R_{0}),\cr
\nonumber \\
b&=&\sin \theta_{2} \sin \theta_{3}(1+R_{0}),\cr \nonumber \\
c&=&\cos \theta_{2}(1+R_{0})-\cos \theta_{1}(1-R_{0}) \nonumber
\end{eqnarray}

In the derivation of the equation(1), it is assumed that the
incident wave $\psi^{in}$ is a plane wave with a wave vector ${\bf
k}$ as $ \psi^{in}(r)=\exp(i {{\bf k} } \cdot  {{\bf r}})$.
\begin{figure}
\epsfxsize=7truecm\epsfbox{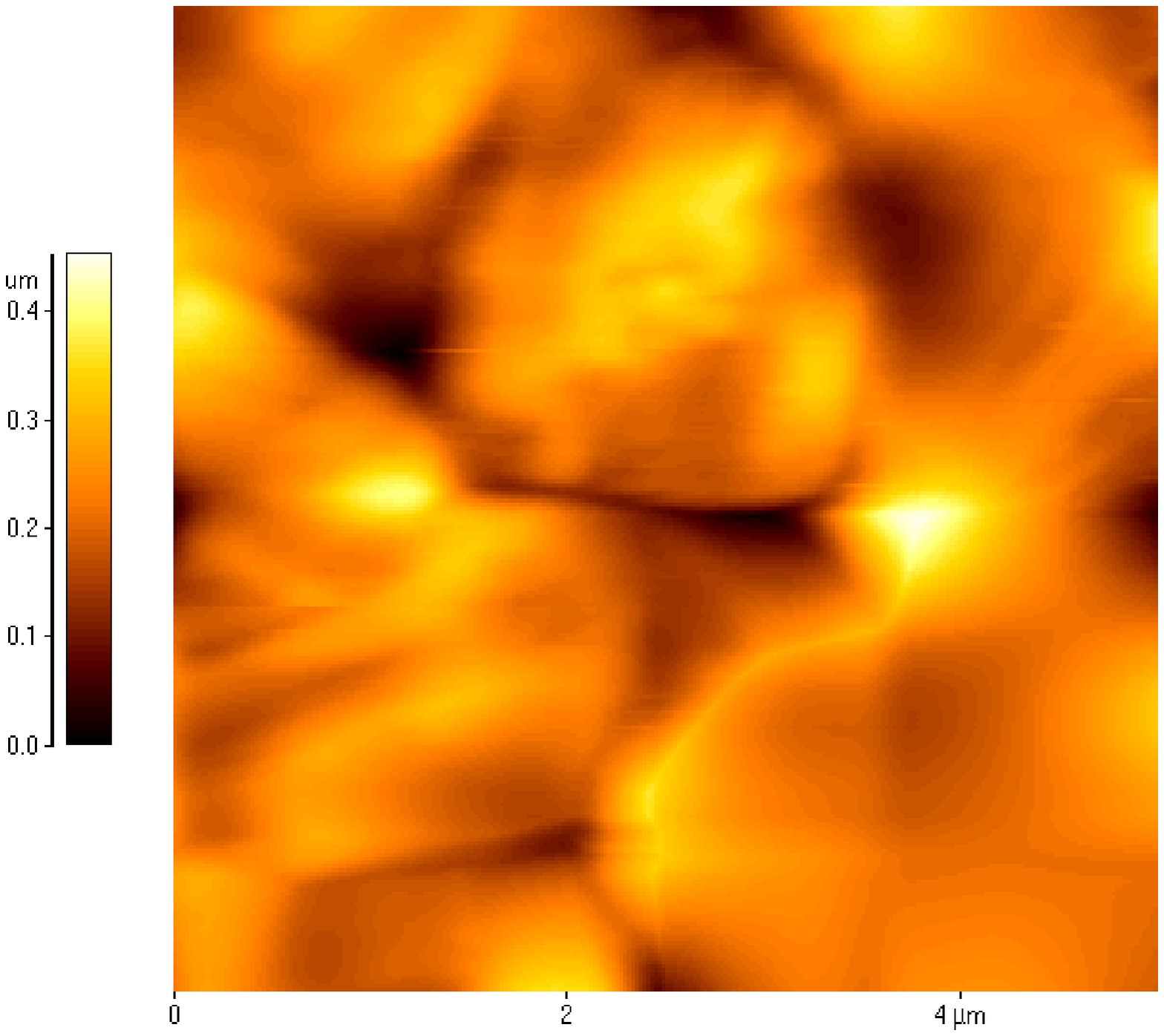}
 \narrowtext \caption{AFM image of the Alumina surface in the length scale
$5 \mu m \times 5 \mu m$ (small scale). }
 \end{figure}
\begin{figure}
\epsfxsize=7truecm\epsfbox{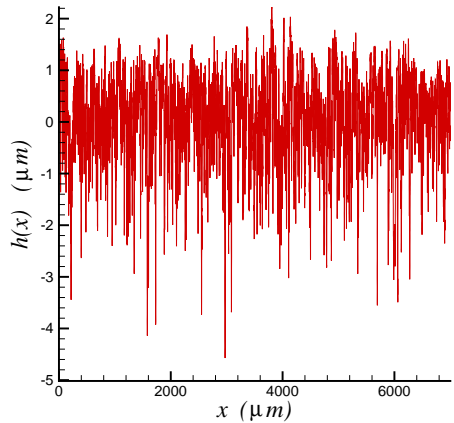}
 \narrowtext \caption{Profilometer scans of the Alumina surface with resolution $0.25
\mu m$ (large scale). }
 \end{figure}

In most cases, the wave scattering models from rough surfaces
implicitly assume that the surface is rough on a single scale.
However, in practice all surfaces are rough on several scales,
ranging from atomic scale to the scale determined by the length of
the surface. Nevertheless, only a finite range of scales are
important in scattering of waves from a surface, i.e. the range
covering the wavelength of the incident radiation.
 Models have
been developed for describing surfaces that consist of high
frequency fluctuations superimposed on a slowly varying roughness
 \cite{Ogilvy}. These models use perturbation theories to describe the
scattering from the high frequency roughness and this is modified
in some manner by the low frequency component \cite{Antoine}. All
of the perturbative methods deal with the effect of the large
scale fluctuations as perturbation to the small scale height
fluctuations. Here, we intend to observe the surface in two scales
with resolutions of nanometer and micrometer. The figure (4),
shows schematically the modulation of small scale height
fluctuations by large scale variations. Various statistical
parameters like the joint height and height gradient PDF, surface
roughness $\sigma$, correlation function $C(R)$, correlation
length $\tau$ etc., were measured in two scale.
\begin{figure}
\epsfxsize=7truecm\epsfbox{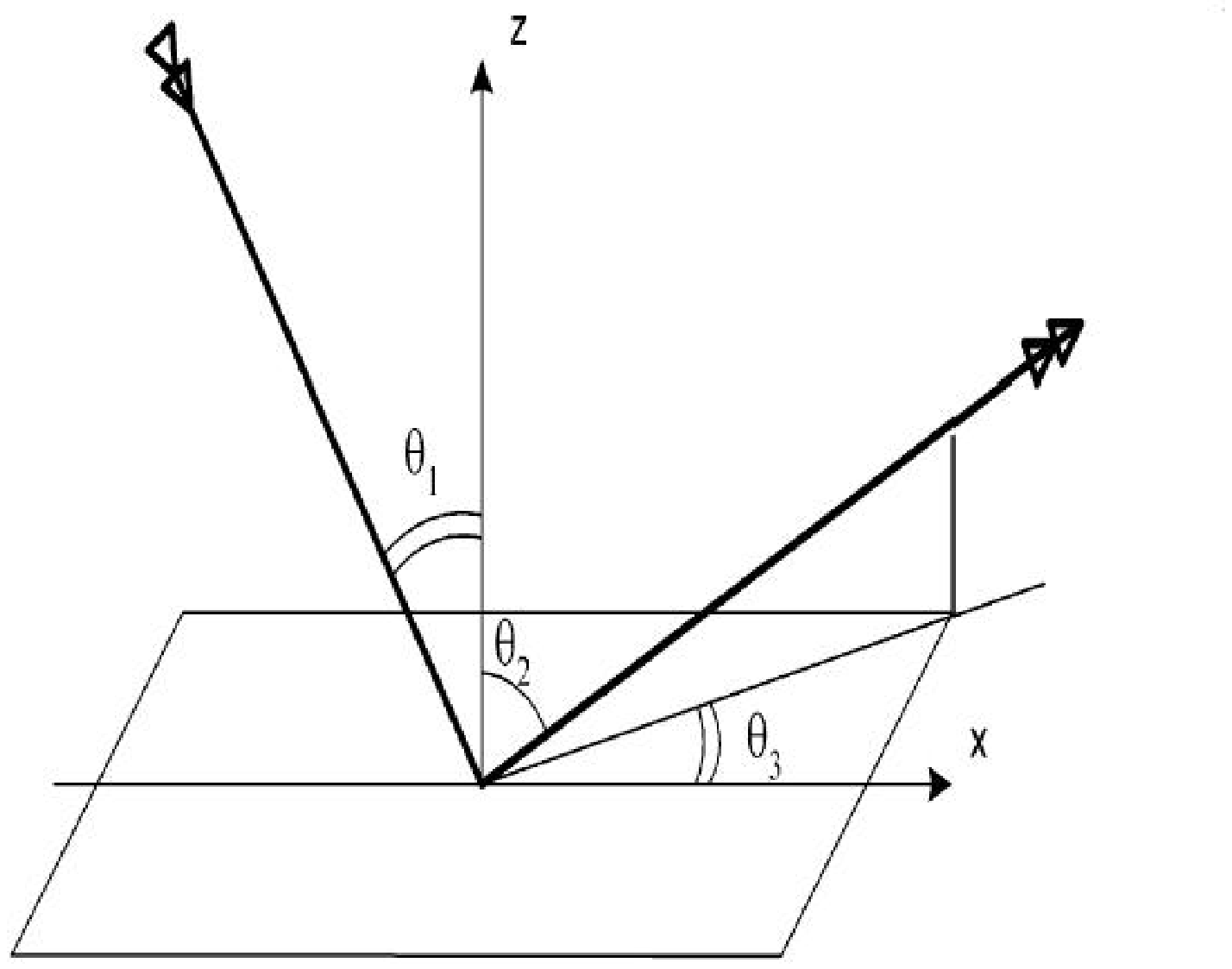}
 \narrowtext \caption{ The geometry of the scattering angles $\theta_{1}, \theta_{2}$
 and $\theta_{3}$.}
 \end{figure}
\begin{figure}
\epsfxsize=8truecm\epsfbox{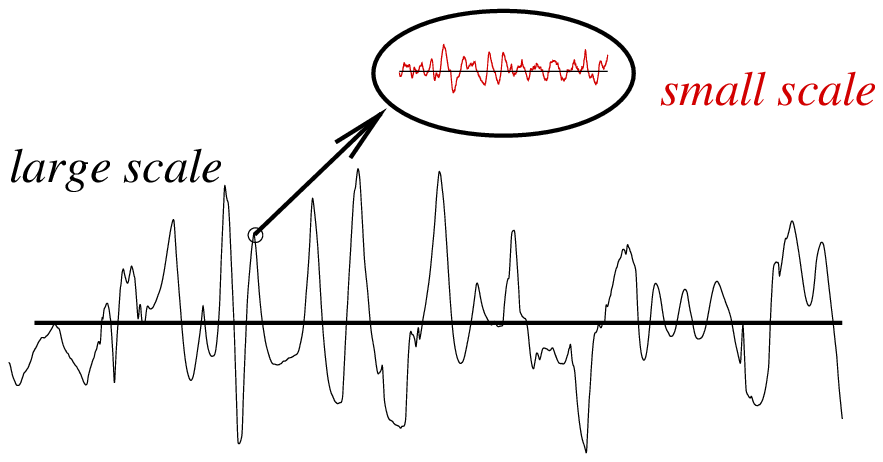}
 \narrowtext \caption{ Two scale observation model of Alumina surface. }
 \end{figure}

In what follows, we are going to describe the non-perturbative two
scale Kirchhoff theory. We first calculate the contributions of
the coherent and the diffuse fields by the Kirchhoff theory in
small scale. The coherent field with a gaussian height
distribution will be  \cite{Ogilvy}:

\begin{equation}
<\psi^{sc}><\psi^{sc}>^{*}=I_{0} \exp(-g)
\end{equation}
where $g=k^{2}\sigma^{2}C^{2}$. Also $k$, $\sigma$ and $I_{0}$ are
the norm of wave vector, surface roughness in small scale and the
scattered reflected intensity of the corresponding smooth surface.
For isotropic surface and for samples with the sizes much larger
than the correlation length $L \gg \tau$, (and for a slightly
rough surface i.e.  $g \ll 1$) the diffuse field
 intensity for Gaussian height distribution will be given by \cite{Ogilvy}:
\begin{eqnarray}\label{11} && <I_{d}> = \cr \nonumber \\
&&\frac{k^{2}F^{2}\tau ^{2}}{4\pi r^{2}}g \exp ({-g}) A_{M}
\exp(-\frac{k^{2}(A^{2}+B^{2})\tau^{2}}{4})
\end{eqnarray}
where, $F=\frac{1}{2}(\frac{Aa}{C}+\frac{Bb}{C}+C)$ and $A_M$ is
the effective area of rough surface which experience the incident
radiation. Therefore, the overall scattered intensity is written
as \cite{Ogilvy}:

\begin{equation}
<I>= I_{0}\exp({-g}) + <I_{d}>
\end{equation}

So far, we have expressed the results of the light scattering from
the surface in small scale. Now we divide the whole surface to
many small pieces (meshes) such that length of which is smallest
scale of our observation. In each mesh we can apply one scale (
small scale) Kirchhoff theory. Therefore for each mesh we have a
similar expression for coherent field as equation (2), but with
different angles which depend on the positions of the small mesh.
In the small scale, we denote the height field in position $x_s$
and $y_s$ with $h_s$. Therefore, one can write the height field in
any position, $\vec{x}$, as follows:
\begin{figure}
\begin{center}
\epsfxsize=7.5truecm\epsfbox{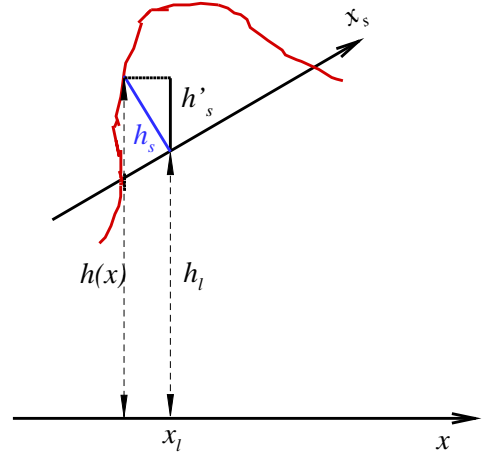} \narrowtext \caption{A two
dimensional scheme, shown the necessary parameters, $h'_s$, here
is the $h_s$ rotated such that it coincide on the vertical time In
the three dimensional case, $h'_s$ is rotated to $h_s$ by Euler
matrices.}
\end{center}
\end{figure}
\begin{eqnarray}
h &&= h_{l} + h'_{s}  \cr \nonumber\\
x &&= x_{l} + x'_{s}  \cr \nonumber\\
y &&= y_{l} + y'_{s}
\end{eqnarray}

The indices $l$ and $s$ denote the large and small scales,
respectively. The vector ( $x'_{s}$, $y'_{s}$ ) is the position of
the $h'_{s}$ on the small scale coordinates. In figure (5) we have
shown the ($h'_s,x'_s,y'_s$) and ($h_s,x_s,y_s$), schematically.
We note that the AFM images will gives us the $h_s(x_s,y_s)$ and
via the large scale topography we will find the $h_l(x_l,y_l)$.
The vectors ($h'_s,x'_s,y'_s$) and ($h_s,x_s,y_s$) can be related
to each other, via rotational Euler matrix, with three rotational
angel $\alpha,\beta$ and $\gamma$ i.e.
$A(\alpha,\beta,\gamma)=R_h(\gamma) R_y(\beta) R_x(\alpha)$.

The Local angles $(\theta_{1},\theta_{2},\theta_{3})$ are defined
by the average plane in the small scale. Therefore, all $a, b, c,
A, B, C$ are constant for all points within the small piece. In
each small scale element $h_{l}$ is fixed so that $\frac{\partial
h_{l}}{\partial x_{s}}=0$. Hence, the total scattered field has
the following expression:
\begin{eqnarray}
\psi^{sc}(r) &=& \sum_{x_{l},y_{l}} [\frac{ik \exp(ikr)}{4 \pi
r}\int \int_{s_{M}}(a_{s}\frac{\partial h'_{s}}{\partial
x'_{s}}+b_{s} \frac{\partial h'_{s}}{\partial y'_{s}} \cr \nonumber\\
&&-c_{s}) \exp{ik(A_{s}x'_{s}+B_{s}y'_{s}+C_{s}(h'_{s}(x'_{s},y'_{s}))} \cr \nonumber\\
&&dx'_{s}dy'_{s}]\exp{ik(A_{l}x_{l} + B_{l}y_{l} + C_{l}h_{l})} \cr \nonumber\\
&=&\sum_{x_{l},y_{l}}\psi_{s}^{sc}(r) \exp ({ik(A_{l}x_{l} +
B_{l}y_{l} + C_{l}h_{l})})
\end{eqnarray}
we note that $\frac{\partial h'}{\partial x'_{s}}= \frac{\partial
h}{\partial x_{s}}$ and the summation is over the small scale
samples modulated by the large scale fluctuations.
 We assume that the joint PDF of heights and its slope of two scales are independent, then the
average of the field scattered in any direction will be given by:
\begin{eqnarray}
&& < \psi_{-e}^{sc}(r) > =  N \sum_{h_{l},\partial_x h_{l}}
\sum_{x_{l},y_{l}} < \psi_{s}^{sc}(r)> \cr \nonumber
\\ &&
\exp (ik C_{l}h_{l})  \exp ( ik (A_{l}x_{l} + B_{l}y_{l}))
P(h_{l},\partial_x h_{l}) \cr \nonumber \\
&=& N \sum_{h_{l},\partial_x h_{l}} < \psi_{s}^{sc}(r)>  \exp
(ik (C_{l}h_{l}))  P(h_{l},\partial_x h_{l}) \cr \nonumber \\
&&\sum_{x_{l},y_{l}} \exp
((ik(A_{l}x_{l} + B_{l}y_{l})))\cr \nonumber \\
&=&N \frac{\sin(kL_{x})}{kL_{x}} \frac{\sin(kL_{y})}{kL_{y}} A_{M}
\cr \nonumber\\
&& \sum_{h_{l},\partial_x h_{l}} <\psi_{s}^{sc}(r)> \exp
(ikC_{l}h_{l})  P(h_{l},\partial_x h_{l})
\end{eqnarray}

The subscript $(-e)$ denotes scattering from the surface without
the edge terms.
 The rough surface has been assumed to be rectangular with extent
$-X \leq x_{0} \leq X, -Y \leq y_{0} \leq Y $. Also $L_{x}$ and
$L_{y}$ are length scales in the scattering area
 (the effective area of light incidence), and $S_M= \frac{\sin(kL_{x})}{kL_{x}}
 \frac{\sin(kL_{y})}{kL_{y}} A_{M}$ is the constant term in all observation
 angles. The quantity $N P(h_{l},\partial_x h_{l})$
 is the number of points with height $h$ and slope $\partial_x h_{l}$.
 It is noted that for a
homogeneous surface $p(h,x)$ is independent of position along the
surface, $x$. In order to do analytical calculation, it is
necessary to assume that the edge effects are non-stochastic, i.e.
$<\psi_{e}> = \psi_{e}$ \cite{Ogilvy}. Based on this assumption,
the coherent part becomes:
\begin{eqnarray}
&&<I_{coh}> =<\psi^{sc}><\psi^{sc}>^{*}= \cr \nonumber \\
&&N^{2} \sum_{h_{1},\partial_x h_{1}, h_{2},\partial_x h_{2}}
|<\psi_{s}^{sc}>|^2 P(h_{1}, \partial_x h_{1})P(h_{2},\partial_x
h_{2})
\cr \nonumber \\
&& \hskip 1cm \exp ( ik(C(h_{2}-h_{1})))
\end{eqnarray}

where $\psi^{sc} = \psi_{e} + \psi_{-e}$. It is noted the
non-stochastic assumption of the edge effect leads to the
cancelation of all terms containing edge effects. In cylindrical
coordinates, for an isotropic surface, the substitutions
$x_{2}-x_{1}=R \cos \theta$ and $y_{2}-y_{1}=R \sin \theta$ can be
made. Since the heights PDF and the heights difference PDF are
independent (we will confirm this assumption in the next section),
i.e. $P(h_{1},\partial_x h_{1}) = P(h_{1}) P(\partial_x h_{1})$.

Define,
\begin{eqnarray}
&&\sum_{h_{1},h_2} dh_1 dh_2 \exp (ik(C(h_{2}-h_{1}))P(h_{1})
P(h_2) \cr \nonumber \\ &=& \chi(kC,-kC,R) \nonumber
\end{eqnarray}
then one finds:
\begin{eqnarray}
<I_{coh}>&=&S_{M}^{2} | \sum_{\partial_x h_{1}} N P(\partial_x
h_{1}) <\psi_{s}^{sc}>|^{2} \cr \nonumber
\\ &&\chi(kC,-kC,R)
\end{eqnarray}
It is known that the total average scattered field in small scale
is $<\psi_{s}^{sc}>=\chi(kC_{s})\psi_{0}^{sc}$.

 For a gaussian
height distribution, the one and two-dimensional characteristic
function is given by:
\begin{eqnarray}
&&\chi(kC_{s})=\frac{1}{\sigma_{s}\sqrt{2\pi}}\int_{-\infty}^{+\infty}
\exp(-\frac{h^{2}}{2\sigma_{s}^{2}})\exp(ikC_{s}h_{s})dh_{s} \cr
\nonumber \\ &&=\exp(-k^{2}C_{s}^{2}\sigma_{s}^{2}/2),
\end{eqnarray}

and

\begin{eqnarray}
 \chi(kC,-kC,R)=\exp(-k^{2}C^{2} \sigma_{l}^{2}(1-C(R))).
\end{eqnarray}
where $C(R)=\frac{<h(r)h(r+R)>}{\sigma_l ^{2}}$, is the surface
correlation function in the large scale.
 Also the average of total intensity are given by :

\begin{eqnarray}\label{I}
&&<I_{tot}> =<\psi^{sc}\psi^{sc^{*}}> \cr \nonumber \\
&&=N \sum_{h_{1},\partial_x h_{1}}
\sum_{x_{0},y_{0}}\sum_{x_{1},y_{1}}P(h_{1},\partial_x
h_{1})<\psi_{s}^{sc}\psi_{s}^{sc^{*}}>
 \cr \nonumber \\
&& \exp (ik( A(x_{2}-x_{1}) + B(y_{2}-y_{1})))
\end{eqnarray}

\begin{figure}
\epsfxsize=7truecm\epsfbox{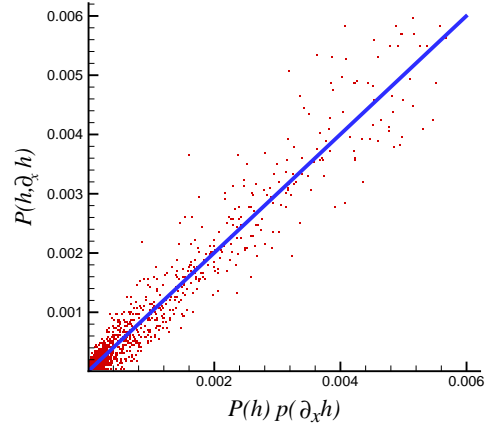}
 \narrowtext \caption{ Joint PDF vs $P(h) P(\partial_{x}h)$, that shows the
height and slope PDFs are almost independent.}
 \end{figure}

Performing the summation we find $\sum_{h_{l}}=N$, where $N$ is
the number of points on the surface.  So, the average total
intensity becomes:

\begin{eqnarray}\label{16}
&&<I_{tot}> =S_{M}^{2}N \sum_{\partial_x h_{1}} N(\partial_x
h_{1})<\psi_{s}^{sc}\psi_{s}^{sc^{*}}>
\end{eqnarray}

Finally, the diffuse field intensity is obtained as:
\begin{eqnarray}\label{17}
 <I_{d}> = <I_{tot}>-<I_{coh}>
\end{eqnarray}

\section{Comparison With Experiments}

Here we test the non-perturbative two--scale Kirshhoff theory with
experiment. For this purpose, we obtain the height profile of the
Alumina sheets as the rough samples, using the
 profilometer in large scale
and the AFM images in small scales. Indeed we intend to observe
the surface in two scales; nano-meter and micron. To use the two
scale theory the surface must possess two conditions.  First, the
PDF of the height and its slope must be independent at small and
large scales, i.e. $P(h_l, \partial_x h_l,h_s, \partial_x h_s)
=P(h_l, \partial_x h_l) P(h_s, \partial_x h_s)$. The homogeneous
rough surfaces possess this condition. Indeed statistical
parameters in small scale ( roughness, exponents, etc.) are
similar at any point of the sample (large scale). This means that
the two PDFs are independent. The second condition is that the
height and height gradient fluctuation must be independent in the
large scale. This means that the joint PDF of the height and
height gradients can be decomposed as $P(h_l, \partial_x h_l) =
P(h_l) P(\partial_x h_l)$. This assumption needs confirmation.  In
figure (6), we have plotted the joint PDF $P(h_{l},\partial_x
h_{l})$ vs $P(h_{l}) P(\partial_x h_{l})$. It is obvious that the
joint PDF vs multiplication of single PDFs fits with a line with
slope one. Considering its statistical error we observe that the
height and height gradient PDFs  are independent. For large values
of $h$ and $\partial_x h$, our assumption becomes poor and thus
uncertainty increases.
\begin{figure}[t]
\epsfxsize=7truecm\epsfbox{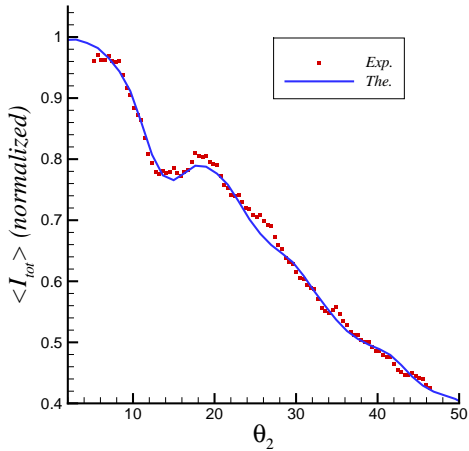}
 \narrowtext \caption{ Comparison of theoretical prediction via two scale Kirshhoff theory and
 experimental results for scattered field (bold symbols).}
 \end{figure}

To compare the experimental observation with those of the
theoretical prediction, we need estimate the several types of PDFs
in small and large scales. In the equation (8), we need to
evaluate the quantity  $<\psi_{s}^{sc}\psi_{s}^{sc^{*}}>$ in the
small scale and PDF of the height gradients in the large scale. To
evaluate the intensity $<\psi_{s}^{sc}\psi_{s}^{sc^{*}}>$, we have
to use the equation (2), where the averaging is done in the small
scale. Therefore we need the PDF of height fluctuation in the
small scale. Also we need other statistical quantities such as
 surface roughness $\sigma$,
correlation function $C(R)$, correlation length $\tau$  etc., in
small and large scales. 
We evaluate the height -height correlation function $< h(x+R) h(x)
>$ vs radial distance $R$ for large scale fluctuations.  We find
the following expressions for the Alumina surface as, $C(R) = 2.14
\exp(-\frac{R^{2}}{608})$ and $1.27 \exp(-0.58R)$, for small and
large scales, respectively. Also the roughness exponent, variance
and scaling length for the small (large) scale have been found as,
$0.85$ ($0.85, 031$), $0.31 \mu m$ ($1.33 \mu m$) and $1.5 \mu m$
($19.4 \mu m$), respectively. It is found that the height PDF in
the two scales are almost gaussian. The estimated statistical
quantities enable us to predict the average total intensity. In
figure (7), we have plotted the experimental observation and
theoretical prediction of total intensity. It is evident that the
theoretical prediction fits with those of experimental
observation. We observe the theory is able to predict small peak
in the angle $\simeq 18^0$ in the variation of the total intensity
vs angle scale $\theta_2$. We note that if one plots the PDF of
height gradient, then  finds that the PDF has also small peaks at
angle scale $\tan^{-1} (\partial_x h) \simeq 9^0$. This means that
the gradient PDF is responsible to have a small peak in the
variation of the total intensity in terms of angle scale (we note
that the slope $\alpha= \tan^{-1}( \partial_x h)$, we produces $2
\alpha$ contribution in the reflection of the light from the
surface). In figure (8), the behavior of the slope PDF
($\partial_{x_l} h_l$) in terms of $\partial_{x_l} h_l$, has been
given. Also as shown in figure (7), the two scale Kirshhoff theory
is able to predict the small peak in the variation of the total
intensity in terms of angle scale. As we observe, there are other
peaks in the figure (7), where the theory can not predict the
peaks for large angle scales. Indeed for these angle scales we
should take into account the shadowing effect \cite{Fuks,Bruce}.
In ref.[12], the validity range of geometrical shadow functions
has been investigated for a randomly rough surface for which the
shadowed Kirchhoff approximation has been shown to give good
results for the scattered intensity distribution. We will discuss
the modification of the two scale Kirshhoff theory by the
shadowing effect elsewhere.
\begin{figure}
\epsfxsize=7truecm\epsfbox{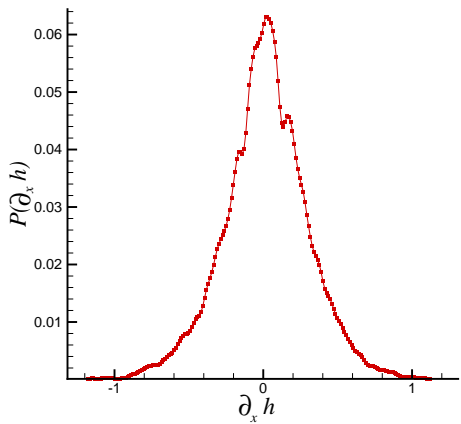}
 \narrowtext \caption{The PDF of height gradient in large scale. }
 \end{figure}

\section{Acknowledgments}
The authors would like to thank the Ministry of science, research
and technology for financial support.



\begin{thebibliography}{99}
\bibitem{Ogilvy} J.A. Ogilvy, Theory of Wave Scattering from Random Rough
Surfaces, Institute of Physics Publishing, Bristol and
Philadelphia, 1991; A. G. Voronovich, Wave scattering from rough
surfaces. Springer-Verlag, Tiergarten-Strasse 17, D-69121
heidelberg, Germany, second updated edition, 1994; T. Elfouhaily
and C. A. Guerin. A critical survey of approximate scattering wave
theories from random rough surfaces, Wave in Random media,
14:R1-R40, 2004.
\bibitem{Kakuen} Kakuen Tang and Richard O. Buckius, Int. J. Heat Mass Transfer
{\bf 14}, 13 (1998).
\bibitem{Nieto} Nieto-Vesperinas, M. and Sunchez-Gil, J. A., Journal of the Optical
Society of America A {\bf 9}, 424-436 (1992).
\bibitem{Joyez}P. Joyez and D. Esteve, Physucal Review B, Volume 64,
155402 (2001)
\bibitem{Dimenna}R. A. Dimenna and R. O. Buckius, ASME Journal of Heat
Transfer, 116,639-645 (1994)
\bibitem{Gu}Z. H. Gu, J. Q. Lu, A. Martinez, E. R. Mendez and A.A.
Maradudin, Optical Letters, 19, 604-606, (1994)
\bibitem{Ingve}Ingve Simonsen, Tamara A. Leskova, and Alexei A.
Maradudin, Physical reviwe B, Volum 63, 245411 (2001)
\bibitem{Calvo}O. Calvo-Perez, J. J. Greffet and A. Sentenac,J. Opt. A: Pure Appl.
Opt. 1 (1999) 560–565. Printed in the UK
\bibitem{Jerome}Jerome Caron, Jacques Lafait and Christine Andraud,
Optics Communications 207 (2002) 17–28
\bibitem{Antoine1}Antoine Soubret and Gerard Berginc, arXiv:
physics/0312133 v1 22 Dec (2003)
\bibitem{Antoine2}Antoine Soubret and Gerard Berginc, arXiv:
physics/0312136 v1 22 Dec (2003)
\bibitem{Bass} Bass F G and Fucks I M 1979 Wave Scattering from
Statistically Rough Surface (Oxford: Pergamon)
\bibitem{Ishimaru} Ishimaru A 1978 Wave propagation and Scattering in
Random Media (New York: Academic)
\bibitem{Vornovich} Vornovich A G 1998 Wave Scattering from Rough
Surfaces (Springer Series on Wave Phenomena vol. 17) (Berlin:
Springer)
\bibitem{Fuks} Iosif M Fuks, Wave In Random Media
{\bf 12}, 401-416 (2002).
\bibitem{Bruce} Neil C Bruce, Wave In Random Media {\bf 14}, 1-12 (2003).
\bibitem{Antoine}Antoine Soubret, Gerard Berginc, and Claude Bourrely,
Physucal Review B, Volume 63, 245411 (2001)
\end{thebibliography}
\end{document}